# Neural Mechanisms of Temporal and Rhythmic Structure Processing in Non-Musicians

**Grigoriy Radchenko[1], Valeriia Demareva[1*], Kirill Gromov[1], Irina Zayceva[1], Artem Rulev[1], Marina Zhukova[1], and Andrey Demarev[1]**

[1]Cyberpsychology Laboratory, Faculty of Social Sciences, Lobachevsky State University, Nizhny Novgorod, Russia

**\* Correspondence:**
Valeriia Demareva
valeriia.demareva@fsn.unn.ru

**Abstract**

Music is increasingly being used as a therapeutic tool in the field of rehabilitation medicine and psychophysiology. One of the main key components of music is its temporal organization. The characteristics of neurocognitive processes during music perception of meter in different tempo variations technique have been studied by using the event-related potentials technique. The study involved 20 volunteers (6 men, the median age of the participants was 23 years). The participants were asked to listen to 4 experimental series that differed in tempo (fast vs. slow) and meter (duple vs. triple). Each series consisted of 625 audio stimuli, 85% of which were organized with a standard metric structure (standard stimulus) while 15% included unexpected accents (deviant stimulus). The results revealed that the type of metric structure influences the detection of the change in stimuli. The analysis showed that the N200 wave occurred significantly faster for stimuli with duple meter and fast tempo and was the slowest for stimuli with triple meter and fast pace.

## 1 Introduction

Music has been an integral part of the life of a human over its documented history. The same holds true for musical ability, i.e., the ability to perceive tone that is inherent to human beings in various degrees. Human beings are capable of perceiving and responding to rhythmic structures from infancy. Studies have shown that the rhythm young children find most comforting and the rate of synchronization with external rhythm change with age: the internal rhythm slows down and the rate of synchronization with the external rhythm decreases (Drake et al., 2000).

The physiological basis for rhythm perception lies in the motor networks, particularly the basal ganglia. Grahn J.A. et al. (2007) found that the basal ganglia and supplementary motor area (SMA) responded with increased activity to rhythm with integer interval ratios and regular accents. At the same time, these areas are active even if there is no movement to the beat of the music (Merchant et al., 2015). Perception is also associated with the interaction of auditory and motor brain areas (Zatorre et al., 2007). Investigating beta-band fluctuations during rhythm perception, Fujioka T. et al (2009) suggested that alongside with the sensorimotor function, the processing of auditory information is also provided by the motor system. Hence, to catch the effect of auditory information on motor system means of monitoring and control are required. Such control is carried out through modeling of upcoming events, which allows not only to anticipate them, but also to save resources for processing incoming information (Schubotz, 2007). Engel A.K. and Fries P. (2010) also found a large network in sensorimotor areas engaged when tracking the beat tempo even without the intention to adjust to the rhythm of movement and concluded that the motor system is activated during passive perception of rhythms.

The process of predicting rhythm events is reflected in event-related potentials (Bouwer & Honing, 2015). There are three components of event-related potentials associated with prediction generation: MMN, N2b, and P3a (Bouwer et al., 2016). In particular, the MMN (mismatch negativity) component reflects the memory system's encoding of information beyond the acoustic properties of sound, and is related to rhythm, melody, and harmony (Näätänen et al., 2007). Therefore, a common research approach is to disrupt the predictions that the listener's brain creates regarding the stimulus being played (Brochard et al., 2003; Honing et al., 2014; Bouwer et al., 2016). This approach involves using different types of stimulus materials for specific studies, such as music-like sounds (Bouwer et al., 2014), rhythms with the same sounds but different time intervals between them (Geiser et al., 2010), stimuli with changing tone of sounds and rhythm (Geiser et al., 2009), monotone isochronous sequences (Schwartze et al., 2011), and 'shaking' sequences with sound skips or changes in tone in unexpected places (Teki et al., 2011).

Music perception can be influenced by various factors. For example, Bouwer and Honing (2015) demonstrated that temporal attending and temporal prediction affect the way a metrical rhythm is processed. They found that intensity increments that fall on the expected beat are detected faster than those off the beat. Brochard et al. (2003) provided physiological evidence for the phenomenon of subjective accentuation, where identical sound events in isochronous sequences are perceived by the listener as unequal. Based on an analysis of the event-related potential components, it has been suggested that perceived disturbances are related to the dynamic unfolding of attention.

Two questions remain unclear regarding music perception: how it is affected by attention, and the listener's musical training. Geiser et al. (2009) showed that musicians are faster and more accurate at detecting errors in rhythm and meter tasks, but the neurophysiological processes that accompany the perception of rhythm and meter did not differ between individuals with and without musical training. However, musical skills are associated with an increased sensitivity to meter, which can be traced to the MMN component at the physiological level (Geiser et al., 2010). Bouwer et al. (2014) found that simple rhythms can be perceived regardless of attention or musical training, and that consistent training improves rhythm perception even without attention directed to musical stimuli. However, differences in rhythm perception between musicians and non-musicians are only evident when attention is directed to the rhythm (Bouwer et al., 2016). In other words, conscious effort is required to manifest musical abilities or skills.

Geiser E. et al. (2010) considers rhythm and meter perception to be the ability to extract not only a regular pulse but also a hierarchical structure with accents of tones from a sequence of sound stimuli. Thus, the study of neural correlates of meter perception in musicians as a model can be extremely useful for the study of patterns of a higher hierarchical level. It is believed that the study of rhythmic series perception can provide a more detailed insight into the dynamic deployment of attention and the physiological correlates of this process (Schwartze et al., 2011). In addition, recent evidence suggests that there are shared neurobiological resources for processing musical rhythms and speech rhythms, which is reflected in reading. Fotidzis T. et al. (2018) applied the expected rhythm disturbance model to reading and found a relationship between musical rhythm abilities and reading comprehension skills reflected in EEG measures (Fotidzis et al., 2018). Recent work by Bartolomé-

This is a provisional file, not the final typeset article

Tomás A. et al. (2019) used musical rhythms to examine changes in brain activity accompanying memories evoked by music pieces of varied styles. Preliminary results of this study showed that the power spectral density of alpha, beta, theta, and gamma frequency bands was affected by the presence/absence of memories evoked by music pieces.

The N200 wave is associated with the detection of novelty or mismatch in the presented stimuli. It is assumed that the amplitude of this wave reflects the degree of mismatch between the processed event and the representation in sensory memory of the 'expected' event (Folstein, Van Petten, 2008). In our previous studies, we showed that the amplitude of the N200 wave is related to the processing of tonal harmony (the relationship between tones in musical sounds), which along with rhythmic organization refers to the key elements of musical syntax in traditional Western music (Radchenko et al., 2019). This suggests that the N200 wave will also be sensitive to changes in temporal organization, reflecting its representation in sensory auditory memory.

The primary goals of the current study are:

1) Comparison of amplitude-temporal characteristics of wave N200 event-related potentials (ERP) when listening to musical fragments with changing rhythmic structure (duple/triple meter);

2) Comparison of amplitude-temporal characteristics of the N200 wave of the ERP while listening to musical fragments with changing rhythmic structure in the context of fast/slow tempo of playback.

The general goal of this study is to investigate how the brain processes the rhythmic structure of music in relation to the tempo (fast vs. slow) and meter (duple vs. triple) factors.

## 2  Materials and Methods

### 2.1  Study sample

20 participants took part in the study (6 men) aged 18-42 years with a median age of 23 years. None of them was a professional musician and had no special musical education other than secondary school. All participants gave written voluntary consent to participate in the study.

### 2.2  Stimuli

Audio signals with the following characteristics were generated to assess neural processing characteristics of the metrical structure and tempo. The rhythmic structure included changes in dynamics (volume) for strong and weak beats. A synthesized complex tone was used as the sample signal. The tone duration was 100 ms, the frequency was 1046 Hz, corresponding to C6 (two octaves above middle C). The sampling rate was 44.1 kHz. The stimuli were recorded using the Ableton live 10 sequencer and Addictive Drums 2 plug-ins. The 'downbeat' and 'upbeat' part was realized through a similar complex tone, but with the volume lowered by 10 dB. A piano preset (Native Instruments Kontakt 5 plugin and The Giant library) was used in the tone coloring.

Using the received tones, 4 variations of rhythmic sequences were composed: duple meter in slow and fast tempo, triple meter in slow and fast tempo. For each variation, a standard version and a deviant version were implemented. The standard version was a successive alternation of a strong and weak beats. In the deviant version, all the weak beats were replaced by strong ones. Thus, the deviant patterns changed the accent pattern of strong and weak beats. The beat durations, depending on the type of rhythmic sequences, were 500 ms in the fast tempo and 900 ms in the slow tempo (Fig. 1).

## 2.3 Study Design

The study consisted of 4 experimental series that differed in tempo (fast and slow) or meter (duple and triple). The order in which the series were presented was random for each participant. A break was taken between sessions to allow the participant to rest.

Each experimental series consisted of 625 audio stimuli, 85% of which were with a standard metric structure (standard stimulus) and 15% with a modified metric structure (deviant stimulus). The order in which the stimuli were presented was random, with at least 4 standard stimuli presented between 2 deviant stimuli and at least 10 standard stimuli presented at the beginning of the procedure.

The study of the temporal characteristics of musical signals is critical to any possible micro-delays between notes, so to eliminate these delays, a single sound file consisting of 625 stimuli was generated and subsequently played. Additionally, events were filled in the generated file, containing the playback time of each stimulus in the file. A cable connecting the COM port of the computer from which the sound file was played, and the peripheral port of the electroencephalograph was used to synchronize the playback time of the stimuli and the EEG recording.

The whole procedure of basic stimulus randomization, audio file generation, audio file playback, and event synchronization via COM port was performed in Presentation software (www.neurobs.com).

All stimuli were played from a separate presentation computer running Presentation software, using a Steinberg UR-12 audio interface through Sennheiser HD 569 headphones at 80 dB. The subjects were seated in a comfortable chair with adjustable backrests and armrests and viewed a video sequence of natural landscapes without sound.

The experimental procedure consisted of the following stages:

- Instruction of the subject;

- Listening to 4 experimental series, with a break between them for rest.

Recording was made using electroencephalograph-analyzer EEGA-21-26 'Encephalan-103' (Medikom-MTD, Taganrog). Recording was performed from 19 electrodes (F7, F3, Fz, F4, F8, T3, C3, Cz, C4, T4, T5, P3, Pz, P4, T6, Fp1 Fp2, O1, and O2) with the conventional 10-20 placement, with reference electrodes on the mastoids and a grounding electrode in the vertex and a signal sampling frequency of 250 Hz. The original signal was filtered with a bandpass filter at 0.5-70 Hz and a notch filter at 50 Hz. The subelectrode impedance during the recording did not exceed 10 kΩ.

## 2.4 Data Analysis

EEG data were processed using Brainstorm software version 20.10.2021 (http://neuroimage.usc.edu/brainstorm, (Tadel et al. 2011)). Initially the signal was converted into EDF format and imported into Brainstorm software. A high-pass filter with a cutoff frequency of 1 Hz and a low-pass filter with a cutoff frequency of 40 Hz were used. Electrodes were referenced to the average reference electrode, calculated as the average value between electrodes A1 and A2. Thus, we used A1 and A2 mastoid for reference channels. Epochs at the time of standard and deviant stimuli with a time window of [-100; 500] ms were cut out for analysis. Each epoch was normalized with its baseline in the [-50; 0] ms window to remove the constant amplitude bias. To eliminate the effect of the artifacts caused by eye movements,
the EOG channels were analyzed by a 200-ms sliding window, and samples with a standard deviation above 35 mV and amplitude above 100 mV were excluded from further analysis. On average, the number of excluded epochs was 15%. To reduce error of the first kind, the difference between standard and deviant stimuli was used for statistical analysis for the Cz electrode, as the one least affected by noise and having a pronounced effect for MMN waves (Luck, Gaspelin, 2017). For this signal, weighted mean amplitude and fractional area latency values were calculated in the range of 150-250 ms from stimulus presentation. The weighted mean amplitude was calculated as the average value of the amplitude in the analysis window. The fractional area latency means the time point that divides the area under the curve into two specified parts (in this case, 50% was used) in the corresponding time range. For statistical analysis, analysis of variance (ANOVA) was used with repeated measures taking into account the factors of tempo (fast vs. slow) and meter (duple vs. triple). Additionally, Tukey's correction for multiple comparisons was applied. Statistical analysis was performed in RStudio (v. 2022.07.2 Build 576). The figures were plotted in Jupyter Notebook (desktop graphical user interface Anaconda Navigator 1.9.12).

## 3 Results

Statistically significant fractional area latency effects were found for the meter factor ($F(1, 19) = 7.496$, $p = 0.013$, partial $\eta^2 = 0.283$). No significant effects were observed for the tempo factor and for the interaction effect of meter and tempo (see table 1). The N200 wave occurred significantly faster for stimuli with duple meter and fast tempo, and the slowest for stimuli with triple meter and fast pace. Analysis of the weighted mean amplitude values showed a significant effect for the meter factor ($F(1,19) = 8.090$, $p = 0.010$, partial $\eta^2 = 0.299$), no significant effects were observed for the tempo factor and for the interaction effect of meter and tempo (see table 1). We can note a tendency that the amplitude of the N200 fast tempo is higher than that for the slow tempo. At the same time, it is also higher for triple meter more than for duple meter. The N200 waves averaged over all participants for standard, deviant stimuli, and their differences are shown in Figure 2. Mean values and confidence intervals of latencies and amplitudes for each category of stimuli are shown in Table 2.

**Table 1 near here**

**Figure 2 near here**

**Table 2 near here**

## 4 General Conclusions

The results revealed that the type of metric structure influences the detection of the change in it, since the N200 waves are larger for the triple meter and occur with a delay. We interpret these results as indicating that under these conditions the disturbances are more prominent and more perceptible, indicating a greater degree of mismatch detected by the nervous system (Winkler et al., 2009). Previous results obtained on a similar type of stimulus for tempo effects (Zhao et al., 2017) did not reach statistical significance in our study. In our work, we did not observe any significant effects for the tempo factor, while our colleagues observed a difference for both the tempo factor and the meter factor. It can be assumed that this effect may be due to the different level of musical experience of the non-musician participants in the two studies. In (Zhao et al, 2017), the total time of private music lessons (less than 2 years) and the time period since the lessons finished end (more than 5 years) are specified as criteria. In our study, none of the participants had a specialized music education. Both criteria are quite general and do not allow for a homogeneous comparison of participants' musical experience. But the general pattern in the distribution of results holds, namely slow tempo reduces the expression of the N200 wave in response to a change in metric structure. This effect may be caused by the need to retain the rhythmic structure pattern in working memory for a longer period of time at a slower tempo. It can also be associated with the fact that for the triple meter the structural disturbance of the weak beat is more perceptible due to the fact that another weak beat was present before it. In order to verify this statement, it is necessary to conduct additional studies for different patterns in triple meter, which would allow for controlling the number of weak beats in the rhythmic pattern and thus estimating how the number of preceding weak beats affects the amplitude of the N200 wave.

The result that slow tempo reduced the amplitude of the N200 wave in response to metric structure change can also be explained by a great prevalence of the meter containing an even number of elements in the beat in the European musical culture, which can smooth the effect of changes in its structure. It is necessary to further assess the level of musicality of the non-musician participants, which will allow the evaluation of their experience and degree of involvement in the interaction with the musical environment. Additionally, the cultural background of the participants should be considered, as individual studies have demonstrated that the differences in cultural backgrounds affect the way people process auditory cues. For example, there is evidence that subjects from tonal language cultures (e.g., Chinese) performed better than subjects from non-tonal language cultures (e.g., French) at distinguishing pitch (Yu et al., 2015).

## 5 Study Limitations

The level of musical training of the participants was not considered by means of the questionnaire. In the future we plan to take this into account. The current study may be limited by controlling for participants' overall level of attention. Although all participants were watching a silent video to reduce blinking and were instructed to ignore the sounds from their headphones, musically trained individuals may unconsciously allocate a different amount of attention resources to sounds than non-musicians.It is necessary to continue the experiment and include professional musicians and children of school age in order to evaluate the age dynamics of wave N200 in the development process.

## 6 Conflict of Interest

The authors declare that the research was conducted in the absence of any commercial or financial relationships that could be construed as a potential conflict of interest.

## 7 Author Contributions

G.R.—experiment design and methodology, data analysis, and preparation of first draft; V.D. —experiment administration, literature review, data processing, and manuscript preparation; I.Z., M.Zh., A.D., and A.R. —assistance with experiment administration and data collection; K.G. —experiment pipeline preparation and data processing. All authors have read and agreed to the published version of the manuscript.

## 8 Funding

This research was funded by the grant of the Federal Academic Leadership Program Priority 2030.

## 12   Data Availability Statement

The data presented in this study are available on request from the corresponding author. The data are not publicly available due to their containing information that could compromise the privacy of research participants.

**Figures' captions**

**Figure 1.** Characteristics of the stimulus material. The used variations of rhythmic sequences in slow (A) and fast (B) tempos with duple (top) and triple (bottom) meters are represented. The color indicates tone replacement in deviant stimuli.

**Figure 2.** Averaged waves of event-related potentials for the 4 stimulus presentation variants (indicated by color). Figure A shows the waves for deviant types of stimuli. Figure B is for standard stimulus types. Figure C represents the difference between the standard and deviant stimulus types.

**Table 1.** Fractional area latency and weighted mean amplitude effects for the meter and tempo factors and their interaction

|  | **Fractional area latency effects** | | | **Weighted mean amplitude effects** | | |
| --- | --- | --- | --- | --- | --- | --- |
| **Effect** | F(1, 19) | p | η2 | F(1, 19) | p | η2 |
| **meter** | 7,496 | 0,013* | 0,283 | 8,090 | 0,010* | 0,299 |
| **tempo** | 0,063 | 0,804 | 0,003 | 0,277 | 0,605 | 0,014 |
| **meter*tempo** | 0,419 | 0,525 | 0,022 | 0,006 | 0,938 | 0,000 |

* - p < 0,05

**Table 2.** Mean values and confidence intervals (CI) of latencies and amplitudes for each category of stimuli

|  |  | Latency [s] |  |  | Amplitude [microvolts] |  |  |
|---|---|---|---|---|---|---|---|
| Stimulus | Meter Tempo | Mean | 95% CI low | 95% CI high | Mean | 95% CI low | 95% CI high |
| Deviant | Duple Fast | 0,201 | 0,193 | 0,209 | 0,902 | 0,326 | 1,477 |
| Standard |  | 0,210 | 0,202 | 0,217 | 0,189 | -0,047 | 0,425 |
| Deviant | Duple Slow | 0,201 | 0,192 | 0,210 | -1,751 | -2,408 | -1,094 |
| Standard |  | 0,201 | 0,194 | 0,208 | -2,264 | -2,886 | -1,642 |
| Deviant | Triple Fast | 0,204 | 0,197 | 0,210 | 1,691 | 0,847 | 2,536 |
| Standard |  | 0,196 | 0,189 | 0,203 | -0,033 | -0,335 | 0,269 |
| Deviant | Triple Slow | 0,200 | 0,195 | 0,206 | 0,715 | 0,090 | 1,340 |
| Standard |  | 0,191 | 0,185 | 0,196 | -0,770 | -1,313 | -0,228 |